\documentclass[twocolumn,showpacs,preprintnumbers,amsmath,amssymb,prl]{revtex4}

\usepackage{graphicx}
\usepackage{dcolumn}
\usepackage{bm}
\usepackage{color}

\begin{document}

\input epsf.sty

\draft
\widetext

\title
{
Doping Dependence of Low-energy Spin Fluctuations in Electron-Doped Cuprates Pr$_{1-x}$LaCe$_x$CuO$_4$
}


\author{M. Fujita$^{1}$}
\email{fujita@imr.tohoku.ac.jp}
\author{M. Matsuda$^{2}$, S.-H. Lee$^{3}$, M. Nakagawa$^{4}$, K. Yamada$^{1}$}

\affiliation{%
$^{1}$Institute for Materials Research, Tohoku University, Sendai, Miyagi 980-0821, Japan\\
$^{2}$Quantum Beam Science Directorate, Japan Atomic Energy Agency, Tokai, Ibaraki 319-1195, Japan\\
$^{3}$Department of Physics, University of Virginia, Charlottesville, Virginia 22904, USA\\
$^{4}$Department of Physics, Tohoku University, Sendai, Miyagi 980-8578, Japan
}

\date{\today}



\begin{abstract}

The low-energy spin fluctuations in the electron-doped Pr$_{1-x}$LaCe$_{x}$CuO$_{4}$ have been investigated over a wide concentration range of 0.07$\leqslant$ $x$ $\leqslant$0.18 that spans from the antiferromagnetic and non-superconducting phase to the superconducting and paramagnetic phase. For all concentrations considered, the low energy excitations exhibit commensurate peaks centered at the ($\pi$, $\pi$) position. Our data show that the characteristics of the excitations, such as the relaxation rate and the overall spectral weight, change rapidly when the system enters the superconducting phase. The spin stiffness also decreases with increasing $x$ in the superconducting phase and is extrapolated to zero at $x$ = 0.21 when the superconductivity disappears. These indicate a close relation between the spin fluctuations and the superconductivity in the electron-doped system.

\end{abstract}



\maketitle

Antiferromagnetism in the doped cuprate Mott-insulator has been extensively investigated due to its rich physics and close relation with the high-$T_{\rm c}$ supercoductivity. In the past two decades, comprehensive studies using spin sensitive probes such as neutron-scattering measurements have clarified the progressive doping evolution of spin correlations in hole-doped ($p$-type) cuprates. In a prototypical $p$-type superconductor La$_{2-x}$Sr$_x$CuO$_4$ (LSCO), the long-range antiferromangetic (AF) order existing in the $x$ = 0 parent compound gets rapidly destroyed upon doping and is replaced by a spin-glass phase at $x$ $\sim$ 0.02 ~\cite{Keimer92}. In the spin-glass phase, an \textit{incommensurate} (IC) spin-density-wave state appears at low temperatures, with a modulation wavevector diagonal to the Cu-O bond direction ~\cite{Wakimoto99,Matsuda00}. Upon further doping, the system becomes superconducting (SC) and the modulation wavevector changes from diagonal to parallel to the Cu-O bond direction~\cite{Fujita02}. The spatial periodicity of the spin-density-wave is inversely proportional to the superconducting transition temperature ($T_{\rm c}$) in the underdoped region~\cite{Yamada98}. When $x$ increases even further into the overdoped region, the low-energy spin fluctuations vanishes that coincide with the disappearance of bulk superconductivity~\cite{Wakimoto04}. These results clearly indicate that the spin correlations are closely related with the superconductivity in the hole-doped superconducting cuprates.

On the other hand, less known is the relation between magnetism and superconductivity in electron-doped ($n$-type) supercondcuting cuprates. Recent neutron-scattering studies revealed the existence of \textit{commensurate} low-energy spin fluctuations centered at ($\pi$, $\pi$) position in both AFM ordered and SC phases of Nd$_{1.85}$Ce$_{0.15}$CuO$_4$ (NCCO)~\cite{Yamada_NCCO} and Pr$_{0.89}$LaCe$_{0.11}$CuO$_4$~\cite{Fujita03_1}. Thus, the nature of the spin fluctuations in cuprates depends on the type of charge carries. Experimental studies on the electron doped cuprates, however, are quite limited so far. Previous neutron-scattering studies of $n$-type cuprates have focused only near the optimally-doped region~\cite{Yamada_NCCO, Fujita03_1, Kang05, Wilson06} and comprehensive investigation on how the spin dynamics evolve with the concentration of the excess electron is still lacking. 

In this letter, we report our neutron scattering measurements on single crystals of Pr$_{1-x}$LaCe$_{x}$CuO$_{4}$ (PLCCO) with several Ce concentrations of 0.07 $\leqslant$ $x$ $\leqslant$ 0.18, spanning the AFM and SC phase transition at $x$ $\sim$ 0.10~\cite{Fujita03_2}. The main results are that the character of the commensurate spin fluctuations do not change much as long as the system is in the AFM phase ($x \leqslant$ 0.11). On the other hand, in the overdoped SC phase ($x \geqslant$ 0.11) , the characteristic relaxation rate, $\Gamma$, and the spin stiffness, $v$, of the spin fluctuations, decrease linearly with the Ce concentration as well as with the superconducting transition temperature, $T_{\rm c}$.  The overall spectral weight of the spin fluctuations decreases with increasing $x$ and, when extrapolated, the spin stiffness goes to zero at $x \sim 0.21$, coinciding with the disappearance of the superconductivity. These behaviors are qualitatively different from the doping dependence of the spin fluctuations observed in the overdoped region of hole-doped cuprates. Their differences and implications to the physics of the two types of SC cuprates are also discussed. 

Single crystals of PLCCO ($x$=0.07, 0.09, 0.11, 0.13, 0.15 and 0.18) were grown by a traveling-solvent floating-zone method. The crystals with dimension typically of $\sim$30 mm in length and 6 mm in diameter were subsequently annealed under Ar gas flow at 920-950 $^{\circ}$C for 10-12 hours~\cite{Fujita03_2}. To determine $T_{\rm c}$, we have measured the superconducting shielding signal on the small portion of the annealed crystals by a superconducting quantum interference device. Furthermore, the average concentrations of Pr, La, Ce and Cu ions were confirmed to be close to nominal values by using an inductively coupled plasma (ICP) spectroscopy. The larger remaining parts of the annealed crystals were used for our neutron scattering experiments. The measurements used were the thermal triple-axis spectrometers TAS-1 and TOPAN at the JRR-3 reactor in the Japan Atomic Energy Agency (JAEA). Energy of final neutrons was fixed to be $E_{\rm f} =$ 14.7 meV, and the horizontal collimations were 80$^{\prime}$-80$^{\prime}$-80$^{\prime}$-180$^{\prime}$ and 50$^{\prime}$-100$^{\prime}$-60$^{\prime}$-180$^{\prime}$ at TAS-1 and TOPAN, respectively. In this paper, the tetragonal \textit{I}4/\textit{mmm} notation was used in which the principal axis in the $ab$ plane are along the Cu-O bond. Typical lattice constants are, for example, $a =$ 3.985 ${\rm \AA}$ and $c =$ 12.32 ${\rm \AA}$ for the optimally-doped $x$=0.11 sample at 3 K. The crystals were mounted in the ($h$ $k$ 0) zone.

\begin{figure}[t]
\begin{center}
\epsfxsize=3.0in\epsfbox{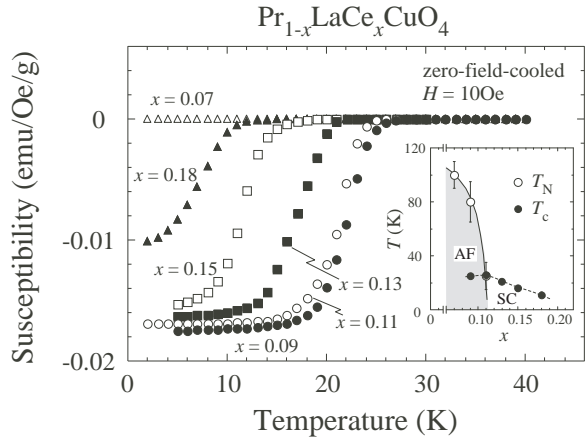}
\caption
{
Temperature dependence of bulk susceptibility obtained from single crystals of Pr$_{1-x}$LaCe$_{x}$CuO$_{4}$ with several different Ce concentrations. The inset 
shows the phase diagram as a function of $x$ and temperature. Here $T_{\rm N}$ (open circles) and $T_{\rm c}$ (filled circles) are the AFM ordering and the SC transition temperature, respectively.
}
\end{center}
\end{figure}

\begin{table}[b]
\caption{\label{tab:fonts} 
Ce concentration $x$, magnetic ordering temperature $T_{\rm N}$ and the average Cu moment $M_{\rm Cu}$, and $T_{\rm c}$ in Pr$_{1-x}$LaCe$_{x}$CuO$_{4}$ determined by inductively coupled plasma, neutron-scattering and SQUID measurements, respectively. 
}
\begin{ruledtabular}
\begin{tabular}{cccccc}
{$x$} & 
{$T_{\rm N}$(K)} & {$M_{\rm Cu}$($\mu$$_{\rm B}$)} & {$T_{\rm c}$(K)} & {Ref.}\\ 
\hline
\hspace{2mm} {0.07} & 
{100(10)} & {0.08(10)} & {-}\\ 
\hspace{2mm} {0.09} & 
{80(15)} & {0.05(5)} & {25(1)} & {[1]}\\ 
\hspace{2mm} {0.11} & 
{25(10)} & {$<$0.01}& {26(1)} & {[2]}\\ 
\hspace{2mm} {0.13} & 
{-} & {-} & {21(1)}\\ 
\hspace{2mm} {0.15} & 
{-} & {-} & {16(2)}  & {[3]}\\ 
\hspace{2mm} {0.18} & 
{-} & {-} & {11(2)}\\ 
\end{tabular}
\end{ruledtabular}
\end{table}

Fig. 1 shows temperature dependence of bulk magnetic susceptibility obtained from several Pr$_{1-x}$LaCe$_{x}$CuO$_{4}$ crystals with $x$ = 0.07, 0.09, 0.11, 0.13, 0.15 and 0.18. The measurements were done under a magnetic field of 10 Oe after zero-field-cooling. Bulk superconductivity appears in all the samples except for $x$ = 0.07. The onset temperature of superconductivity, $T_{\rm c}$, is maximal at 26 K for $x$ = 0.11 and gradually decreases with further increasing $x$. The $x$ = 0.07, 0.09 and 0.11 samples exhibit magnetic order at low temperatures, while no evidence of AFM order was detected for $x$  $\geqslant$ 0.13. In the AFM phase ($x$ $\leqslant$ 0.11), elastic magnetic intensity normalized by the sample volume drastically decreases with increasing $x$. In the vicinity of phase boundary, $x \sim 0.10$, AFM order coexists with superconductivity, as summarized in Table I and in the inset of Fig. 1.

\begin{figure}[t]
\begin{center}
\epsfxsize=3.38in\epsfbox{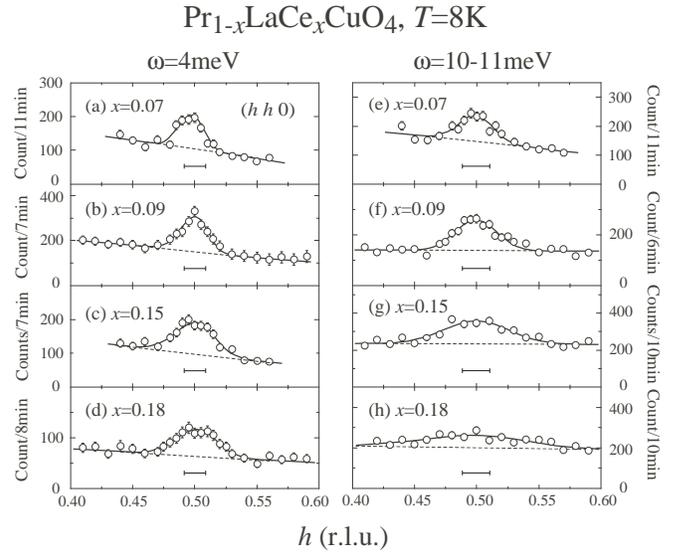}
\caption
{
Constant-$\omega$ scans with $\omega =$ 4 meV ((a)-(d)) and 10-11 meV ((e)-(h)) obtained from Pr$_{1-x}$LaCe$_{x}$CuO$_{4}$ with different $x$; 0.07 ((a), (e)), 0.09 ((b), (f)), 0.15 ((c), (g)) and 0.18 ((d), (h)). Solid lines are fits to a single Gaussian convoluted with the instrumental {\bf Q}-resolution. Dashed lines are background. Horizontal bars represent the the instrumental {\bf Q}-resolution. 
}
\end{center}
\end{figure}

In order to investigate how magnetic fluctuations evolve as Pr$_{1-x}$LaCe$_x$CuO$_4$ changes from insulating to superconducting with increasing $x$, we have performed a series of inelastic neutron scattering measurements on the system with several different $x = $ 0.07 (insulating and AFM ordered), 0.09 , 0.11(superconducting and AFM ordered), 0.13, 0.15, and 0.18 (superconducting and paramagnetic (PM)). Fig. 2 shows some typical data obtained from constant-$\omega$ scans. When the AFM ordering exists ($x$ = 0.07 and 0.09), the low energy spin fluctuations are centered at the characteristic wavevector of the AFM ordering, (1/2,1/2,0), and the peak is almost Q-resolution limited (see Fig. 2 (a),(b)), no matter if the system is superconducting or not. When the AFM ordering disappears upon further doping ($x >$ 0.11), the spin fluctuations remain commensurate but broaden considerably in the momentum space. This broadening is more apparent at higher energies, as shown in Fig. 2 (g),(h), for $\omega =$ 10-11 meV. For quantitative analysis, we have fitted the Q-dependence of the inelastic scattering intensity, $I ({\bf Q},\omega)$, to a single Gaussian, $I(\omega)\exp^{-ln(2)\{({\bf Q}-{\bf Q}_{\rm AFM})/\kappa\}^2}$, convoluted with the instrumental resolution. (We use units with $\hbar$=1.) Here $\kappa$ is the Half-Width-of-the-Half-Maximum (HWHM), the inverse of the dynamic spin correlation length, and $I(\omega)$ is the integrated intensity over Q for a given $\omega$. The solid lines in Fig. 2 are the results of the fit.

\begin{figure}[t]
\begin{center}
\epsfxsize=3.15in\epsfbox{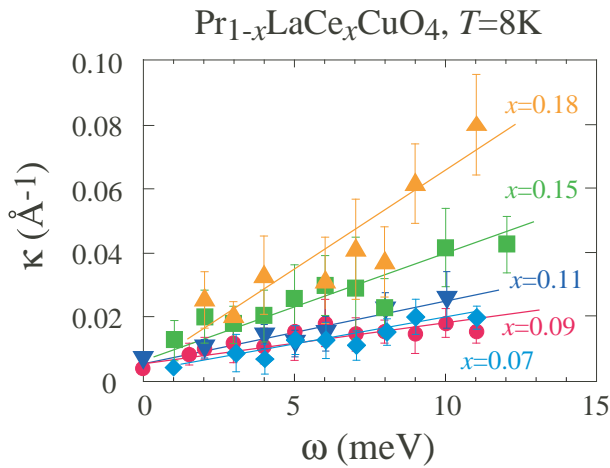}
\caption
{(Color) 
$\omega$-dependence of resolution corrected peak-width (half width at half maximum) $\kappa$ of commensurate peak for Pr$_{1-x}$LaCe$_{x}$CuO$_{4}$ with $x$=0.07, 0.09, 0.11, 0.15 and 0.18.}
\end{center}
\end{figure}

We have performed several constant-$\omega$ scans with various energies from $\omega =$ 1 meV to 12 meV to obtain the $\omega$-dependence of $\kappa$ and that of the Q-integrated intensity. Fig. 3 shows the resulting $\kappa (\omega)$ for several different Ce concentrations. For all concentrations considered, $\kappa$ increases linearly with $\omega$ upto $\sim$12 meV. The slope, $\kappa/\omega$, that corresponds to the inverse of the spin stiffness of spin-wave excitation, $1/v$, however, changes with $x$. When we fit the data to a linear function and obtain the slope, we get $\kappa/\omega$ = 1.5(3) $\times 10^{-3}$ (meV ${\rm \AA}$)$^{-1}$ for $x$ = 0.07 (nonsuperconducting and AFM) and for $x$ = 0.09 (superconducting and AFM), 3.4(7) $\times 10^{-3}$ for $x = 0.15$ (SC and PM) and 6.3(14) $\times 10^{-3}$ for $x = 0.18$ (SC and PM). The value for the AFM phase is very close to the value of 1.6 $\times 10^{-3}$ observed in Pr$_2$CuO$_4$~\cite{Bourges97}. These results confirm that the characteristics of the low energy spin fluctuations does not change in the electron-doped cuprate as long as the AFM order exists, irrespective of the superconductivity. On the other hand,  $\kappa/\omega$ increases abruptly in the overdoped SC and paramagnetic phase.

\begin{figure}[t]
\begin{center}
\epsfxsize=2.95in\epsfbox{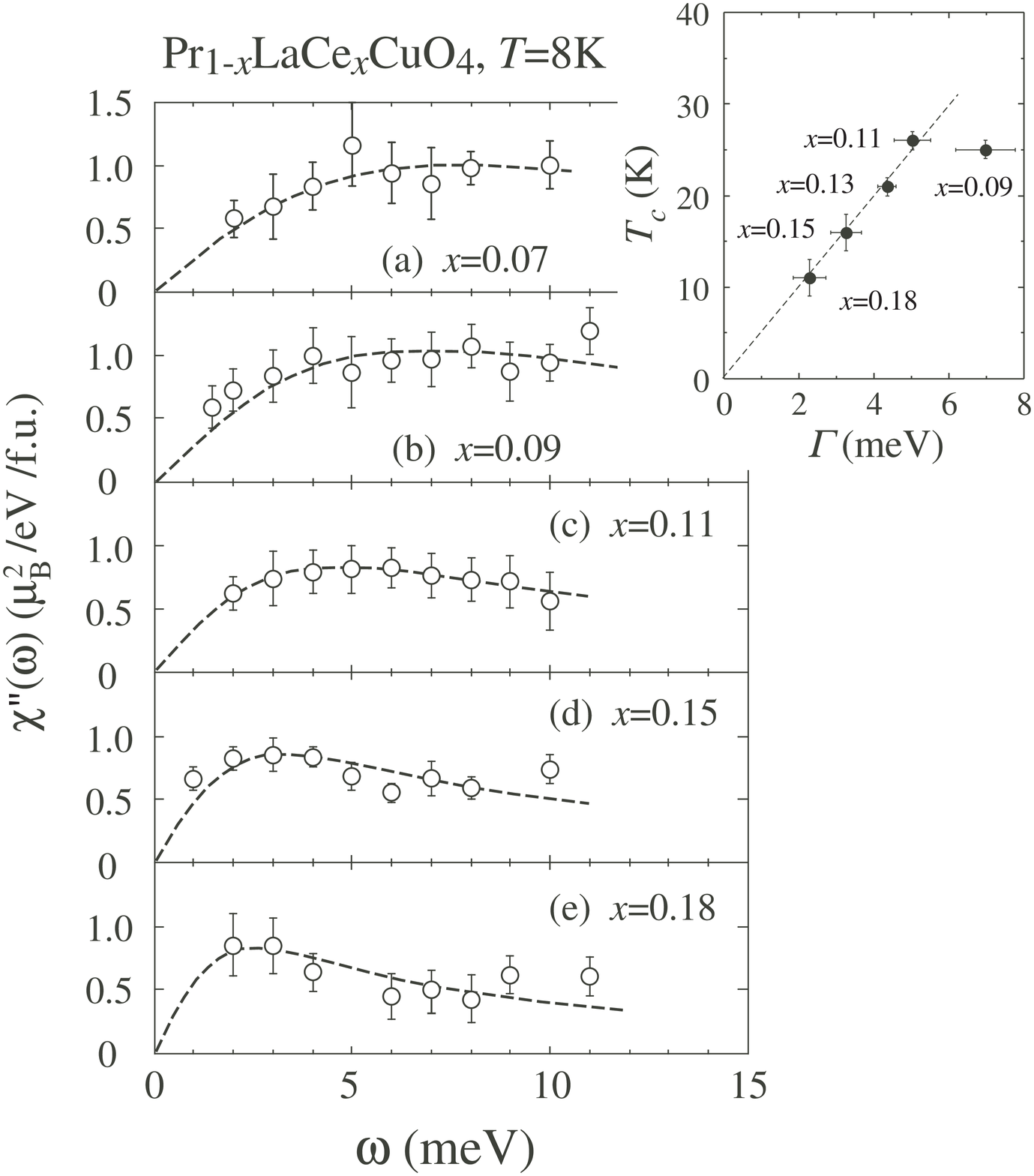}
\caption
{
$\omega$-dependence of local spin susceptibility $\chi^{\prime\prime}$ for Pr$_{1-x}$LaCe$_{x}$CuO$_{4}$ with $x$=(a) 0.07, (b) 0.09, (c) 0.11, (d) 0.15 and (e) 0.18. Dashed lines are fitted results by $\chi^{\prime\prime}\propto\Gamma\omega/(\Gamma^{2}+\omega^{2})$. Inset figure shows the $x$-dependence of $\Gamma$. Inset figure shows the $T_{\rm c}$ as a function of $\Gamma$. Dashed line is the fitted result for $x$$\geqslant$0.11 samples to a linear function. 
}
\end{center}
\end{figure}

Fig. 4 shows that the imaginary part of the dynamic susceptibility, $\chi^{\prime\prime} (\omega)$, as a function of $\omega$ for several Ce concentrations. The $\chi^{\prime\prime} (\omega)$ were obtained by normalizing the Q-integrated intensity $I (\omega)$ to an acoustic phonon around a nuclear (1,1,0) Bragg reflection and by using the detailed balance relation $\chi^{\prime\prime} (\omega) = \pi I (\omega) \cdot (1-exp^{-\omega/k_{\rm B}T})$. In the nonsuperconducting AFM phase ($x = 0.07$), the low energy $\chi^{\prime\prime} (\omega)$ upto $\sim$12 meV gradually increases with $\omega$ and becomes constant for $\omega > 5$ meV. This behavior does not change for $x = 0.09$ and 0.11 in which superconductivity coexists with AFM order. On the other hand, for the paramagnetic SC phase ($x > 0.11$), $\chi^{\prime\prime} (\omega)$ decreases at energies higher than 5 meV, and its spectral weight shifts to lower energies, resulting in shifting of the characteristic energy of the spin fluctuations to lower energies with doping. To extract the relaxation rate, $\Gamma$, we fitted $\chi^{\prime\prime} (\omega)$ to a simple Lorentzian $\chi^{\prime\prime} (\omega) \propto \Gamma \omega/ (\Gamma^2+\omega^2)$. The results of the fit are shown as dashed lines in Fig. 4 and the optimal $\Gamma$ is plotted in the inset.  For $x$ $\geqslant$ 0.11, $\Gamma$ is linearly proportional to the SC transition temperature, $T_{\rm c}$. This is quite different from the behavior observed in the hole-doped cuprates; for instance, in the overdoped region of LSCO, the shape of $\chi^{\prime\prime}$($\omega$) at low energies ramains the same with $\Gamma$ = 6 meV, even though the overall spectral weight decreases. The reduction of $\Gamma$ and the peak-broadening in momentum upon doping observed in the PLCCO are consistent with theoretical prediction based on the $t$-$J$ model~\cite{Tohyama}.

\begin{figure}[t]
\begin{center}
\epsfxsize=2.4in\epsfbox{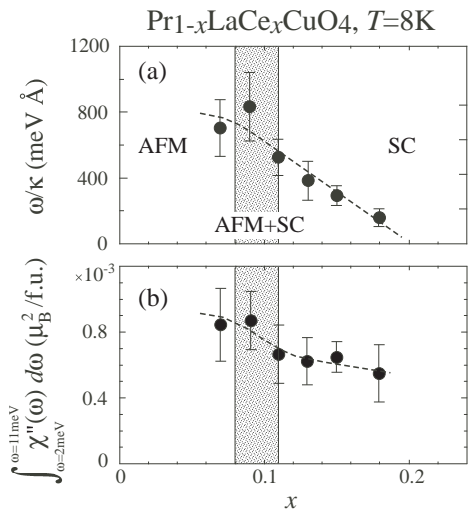}
\caption
{
 {Doping dependence of the low energy spin fluctuations: (a) the spin stiffness, $\omega/\kappa$, and (b) the partial spectral weight obtained by integrating $\chi{\prime\prime} (\omega)$ from 2 meV to 11 meV, as a function of $x$. Dashed lines are guides to the eye.}
}
\end{center}
\end{figure}

The different behaviors in the the spin fluctuations between hole-doped and electron-doped SC cuprates can be understood in the following way. In the case of hole-doped cuprates, it is increasingly evident that there exists a phase separation between SC and normal Fermi liquid phases\cite{Uemura04,Tanabe05}. The volume fraction of the SC region decreases with increasing doping concentration, which coincides with the reduction of the spin fluctuations. This suggests that the IC spin fluctuations in hole-doped cuprates comes from the SC region separated from the normal metallic phase. On the other hand, in the case of electron-doped PLCCO, the the overall spectral weight does not change much with doping concentration(See Fig. 5 (b) showing the $x$-dependence of $\omega$-integrated $\chi$$^{\prime\prime}$), but the characteristic $\Gamma$ of the spin fluctuations strongly depends on $x$. These can be explained not by such a phase separation model. These different behaviors come from the fact that in the electron-doped PLCCO the doped excess electrons go randomly into the CuO planes, inducing a random distribution of nonmagnetic Cu$^{1+}$ (3d$^{10}$) ions and resulting in a system with quenched random site dilution \cite{Mang04}. As consequences, upon doping the commensurate spin fluctuations get broad in Q-space and its intensity decreases. The spin stiffness, $v = \omega/\kappa$, also decreases, which can be understood as a result of reduction of average coupling constants per a magnetic Cu$^{2+}$ ion. These changes are, however, not linear over the entire region of Ce concentration spanning from the non-superconducting AFM phase to the superconducting paramagnetic phase. The low energy spin fluctuations are not so sensitive to the site dilution in the AFM phases, but become sensitive when the system enters the superconducting and paramagnetic phase. 
This doping dependence of spin fluctuation might be related with the change in mobility of doped electrons on crossing the phase boundary. 
Our results show that in the SC and PM phase, $\omega$/$\kappa$ and $\Gamma$ decrease with $x$ and both can be extrapolated to zero at $x$ $\sim$ 0.21 (See Fig. 5(a) for $x$-dependence of $\omega$/$\kappa$), at around which the superconductivity vanishes. The coincidence of the reduction of spin correlations and disappearance of SC phase indicates a close relation between the spin fluctuations and the superconductivity in this electron-doped cuprate. It is to be emphasized that the critical value of $x = 0.21$ is well below the critical value of dilution, $\sim$ 0.41\cite{Newman00}, to destroy percolation in the two-dimensional square spin system and our observed behaviors cannot be easily explained by a simple model of randomly diluted quantum spin systems. Understanding the intimate relation between the spin fluctuations and the superconductivity requires further experimental and theoretical studies on the electron-doped cuprates .

We thank K. Ishida, R. Kadono, T. Tohyama, H. Yamase, S. Wakimoto and G-q. Zheng for their valuable discussions. We acknowledge M. Sakurai for the support on crystal growth at Tohoku University, and M. Kofu and K. Hirota for the technical assistance of ICP measurements at University of Tokyo. This work was supported in part by the Japanese Ministry of Education, Culture, Sports, Science and Technology, Grant-in-Aid for Scientific Research, for Encouragement of Young Scientists (A), 10304026, 2005.


\begin{references}
\bibitem{Keimer92} B. Keimer, A. Aharony, A. Auerbach, R.J. Birgeneau, A. Cassanho, Y. Endoh, R.W. Erwin, M.A. Kastner, G. Shirane, Phys. Rev. B \textbf{45}, 7430 (1992).
\bibitem{Wakimoto99} S. Wakimoto, G. Shirane, Y. Endoh, K. Hirota, S. Ueki, K. Yamada, R.J. Birgeneau, M.A. Kastner, Y.S. Lee, P.M. Gehring, and S.H. Lee, Phys. Rev. B \textbf{60}, R769 (1999).
\bibitem{Matsuda00} M. Matsuda, M. Fujita, K. Yamada, R.J. Birgeneau, M.A. Kastner, H. Hiraka, Y. Endoh, S. Wakimoto, and G. Shirane, Phys. Rev. B \textbf{62}, 9148 (2000).
\bibitem{Fujita02} M. Fujita, K. Yamada, H. Hiraka, P.M. Gehring, S.H. Lee, S. Wakimoto, and G. Shirane, Phys. Rev. B \textbf{65}, 064505 (2002). 
\bibitem{Yamada98} K. Yamada, C.H. Lee, K. Kurahashi, J. Wada, S. Wakimoto, S. Ueki, H. Kimura, Y. Endoh, S. Hosoya, G. Shirane, R.J. Birgeneau, M. Greven, M.A. Kastner, and Y.J. Kim, Phys. Rev. B \textbf{57}, 6165 (1998).
\bibitem{Wakimoto04} S. Wakimoto, H. Zhang, K. Yamada, I. Swainson, Hyunkyung Kim, and R.J. Birgeneau, Phys. Rev. Lett. \textbf{92}, 217004 (2004). 
\bibitem{Yamada_NCCO} K. Yamada, K. Kurahashi, T. Uefuji, M. Fujita, S. Park, S. H. Lee, and Y. Endoh, Phys. Rev. Lett. \textbf{90}, 137004 (2003).
\bibitem{Fujita03_1}  M. Fujita, S. Kuroshima, M. Matsuda, and K. Yamada, Physica C \textbf{392-396}, 130 (2003).
\bibitem{Kang05} H.J Kang, P. Dai, H.A. Mook, D.N. Argyriou, V. Sikolenko, J.W. Lynn, Y. Kurita, S. Komiya, and Y. Ando, Phys. Rev. B \textbf{71}, 214512 (2005).
\bibitem{Wilson06} Stephen D. Wilson, Shiliang Li, Pengcheng Dai, Wei Bao, Jae-Ho Chung, H. J. Kang, Seung-Hun Lee, Seiki Komiya, Yoichi Ando, and Qimiao Si, Phys. Rev. B \textbf{74}, 144514 (2006).
\bibitem{Fujita03_2}  M. Fujita, T. Kubo, S. Kuroshima, T. Uefuji, K. Kawashima, K. Yamada, I. Watanabe, K. Nagamine, Phys. Rev. B \textbf{67}, 014514 (2003).
\bibitem{Bourges97} P. Bourges, H. Casalta, A.S. Ivanov, and D. Petitgrand, Phys. Rev. Lett. \textbf{79}, 4906 (1997). 
\bibitem{Tohyama} T. Tohyama, Phys. Rev. B \textbf{70}, 174517 (2004).
\bibitem{Mang04} P.K. Mang, O.P. Vajk, A. Arvanitaki, J.W. Lynn, and M. Greven, Phys. Rev. Lett. {\bf 93}, 027002 (2004). 
\bibitem{Uemura04} Y.J. Uemura, Solid State Commun. \textbf{126}, 23 (2003).
\bibitem{Tanabe05} Y. Tanabe, T. Adachi, T. Noji, Y. Koike, J. Phys. Soc. Jpn. \textbf{74}, 2893 (2005).
\bibitem{Newman00} M. E. J. Newman and R. M. Ziff, Phys. Rev. Lett. \textbf{85}, 4104 (2000).
%
\end{references}
\end{document}